**Title:** An informed thought experiment exploring the potential for a paradigm shift in aquatic food production

**Authors:** Caitlin D. Kuempel[1,2*], Halley E. Froehlich[1,3,4], and Benjamin S. Halpern[1,5]

**Affiliations:**

[1]National Center for Ecological Analysis and Synthesis, University of California, Santa Barbara, 735 State Street, Suite 300, Santa Barbara, CA 93101, USA

[2] School of Biological Sciences, University of Queensland, St. Lucia, QLD 4072, Australia

[3]Environmental Studies, University of California, Santa Barbara, Santa Barbara, CA 93106

[4]Ecology, Evolution, and Marine Biology, University of California, Santa Barbara, Santa Barbara, CA 93106

[5]Bren School of Environmental Science and Management, University of California, Santa Barbara, Santa Barbara, CA 93106, USA

*Correspondence to:
Caitlin D. Kuempel
+61402284982
c.kuempel@uq.edu.au
School of Biological Sciences
University of Queensland
St. Lucia, QLD, 4072

**Abstract**

The *Neolithic Revolution* began approximately 10,000 years ago and is characterized by the ultimate, nearly complete transition from hunting and gathering to agricultural food production on land. The Neolithic Revolution is thought to have been catalyzed by a combination of local population pressure, cultural diffusion, property rights and climate change. We undertake a thought experiment that examines trends in these key hypothesized catalysts of the Neolithic Revolution and patterns of today to explore whether society could be on a path towards another paradigm shift in food production: away from hunting of wild fish towards a transition to mostly fish farming. We find similar environmental and cultural pressures have driven the rapid rise of aquaculture, during a period that has now been coined the *Blue Revolution,* providing impetus for such a transition in coming decades to centuries (as opposed to millennia). However, we also highlight the interacting and often mutually reinforcing impacts of 1) technological and scientific advancements, 2) environmental awareness and collective action and 3) globalization and trade influencing the trajectory and momentum of the Blue Revolution from patterns and processes of the Neolithic Revolution. We present two qualitative narratives that broadly fall within two future trajectories of seafood production: 1) a ubiquitous aquaculture transition and 2) commercial aquaculture and fisheries coexistence. Each narrative contains two sub-narratives based on differing management and regulatory strategies for aquaculture and fisheries. This scenarios approach aims to encourage logical, forward thinking, and innovative solutions to complex systems' dynamics. Scenario-based thought experiments are useful to explore large scale questions, increase the accessibility to a wider readership, and ideally catalyze discussion around proactive governance mechanisms. We argue the future is not fixed and society now has greater foresight and capacity to choose the workable balance between fisheries and aquaculture that supports economic, environmental, cultural and social objectives through combined planning, policies, and management.

**Introduction**

Producing enough food for the ever-growing human population while staying within planetary boundaries is one of the greatest challenges facing humanity (Springmann et al., 2018). Aquaculture, or aquatic farming, has the potential to play a key role in meeting this challenge due to its immense capacity for growth, particularly in the marine sector (Costello et al., 2020a; Gentry et al., 2017a). The rise of aquaculture over the past several decades has been coined the "*Blue Revolution*" and is considered a new era in food production. However, there is some justifiable concern around aquaculture growth, and it is not yet clear what this new dominant seafood source means for the value of and dependence on wild caught seafood and the health and sustainability of aquatic environments (Mccauley et al., 2015; Naylor et al., 2000) (Fig. 1).

Researchers have theorized about the potential for the Blue Revolution to lead to a Blue Transition, or a more complete shift away from commercial fisheries to aquaculture, in coming centuries (Nahuelhual et al., 2019). Indeed, it has been argued fisheries and aquaculture lie on a continuum of aquatic production (Anderson, 2002; Anderson et al., 2019). This would also not be the first time that humankind has been faced with such a critical paradigm shift in food production. The transition from hunting and gathering to farming on land (i.e., the Neolithic Revolution), began nearly 10,000 years ago and is theorized to have been spurred by a combination of social, environmental and cultural pressures (e.g., local population pressure, cultural diffusion, climate change, property rights; Table 1). Now, several millennia later, land-based hunting and gathering largely exist for subsistence and recreational purposes, while agriculture is credited as the catalyst to modern civilization (Costa-Pierce, 2002; Weisdorf, 2005) and dominates global landscapes (Venter et al., 2016).

Here we reflect on this past terrestrial transition in the context of current patterns in aquatic food production. We highlight the main drivers hypothesized in primary literature to have contributed to the shift to agriculture across time and space, which we find are also identified as contributing factors to the Blue Revolution, albeit at more global and immediate scales (Section 1). The broader, parallel patterns of the two food systems provide important context around the idea that a more ubiquitous transition from fishing to aquaculture is at least possible, especially in the long term (i.e., centuries). However, we also discuss the shifted baseline of human advancements over the past 10,000 years, which inherently challenges and precludes exact predictions of the trajectory, rate of change, potential environmental impacts, and role of both aquaculture and fisheries within society in the future (Section 2, Table 1). In other words, the future is not fixed. We reflect on the quantitative land-based and aquatic food systems patterns – similarities (Section 1) and differences (Section 2) – to form a strong scientific foundation that underpins a qualitative, exploratory scenario approach (Reilly and Willenbockel, 2010). Our scenarios approach uses the standing evidence to explore the potential of a paradigm shift from fishing to aquatic farming (Section 3). Such an approach is common throughout scientific publications in the food systems and conservation literature (e.g., Reilly and Willenbockel 2010, Evans et al., 2013; Gephart et al., 2020; Louder and Wyborn, 2020; Merrie et al., 2018; Peterson et al., 2003; Reilly and Willenbockel, 2010) An exploratory scenario approach is particularly useful for large scale problems with high levels of complexity and uncertainty (Reilly and Willenbockel 2010), which can increase the accessibility and relevance of these concepts to a wider readership beyond academic disciplines (Downs, 2014; Ogilvy et al., 2014; Swart et al., 2004). Specifically, our approach can be classified as an informed thought experiment: hypothesizing future outcomes of complex problems based on existing patterns and data (Aligica, 2005). The goal of this exercise is thus to explore the future balance of wild-caught fisheries and aquaculture through a new lens, and to encourage high-level discussion to lead to science-based, forward thinking, and innovative solutions, rather than a prescriptive outcome or projection.

**Catalysts of domesticated food production on land and in the sea**

*Population pressure & efficiencies of food*

One of the leading theories for the onset of the Neolithic Revolution is centered around local overpopulation events about 15,000 years ago and limits on wild resources (Cohen, 1977; Weisdorf, 2005). It is difficult to measure prehistoric populations, but around this time all inhabitable areas of Europe, Asia and the Americas had been colonized (Cohen, 1977), and research suggests that human population in eastern North America nearly doubled (Weitzel and Codding, 2016). A shift from nomadic to sedentary communities is thought to have led to population pressure, extinction of large megafauna and overexploitation of local resources by early human hunters ("Overkill hypothesis",(Weisdorf, 2005)) that made hunting and gathering inefficient, and the shift to agriculture more viable and perhaps inevitable as a result.

Today, global population has now surpassed 7.6 billion and is projected to reach ca. 10 billion by mid-century (UN DESA, 2017). Population growth, combined with impacts of migration, globalization and increased affluence and consumption, continue to significantly deplete natural resources, drive species extinctions and create serious concerns over our ability to feed the world while staying within planetary limits (Springmann et al., 2018). The sea has been proposed as a leading resource to meet growing protein demand, with the potential for added benefits for environmental and human health (Tilman and Clark, 2014; Willett et al., 2019). In fact, average annual growth in seafood consumption (marine and freshwater, 3.2%) outpaced population growth (1.6%) and apparent meat consumption from all terrestrial animals combined (2.8%) (FAO, 2018).

In line with the theory of growing demand versus cost to obtain food, currently, commercial fisheries can no longer meet seafood demand, nor projected global demands of 215 Mt that are expected in coming decades (Béné et al., 2015; Rice and Garcia, 2011). Global seafood supply now depends on the success and persistence of aquaculture, which has clearly shifted from an "emerging" to dominant seafood source over a comparatively short period of time, accounting for ca. 50% of global seafood production (freshwater and marine) and growing (FAO, 2018). Fisheries' resource scarcity has made some commercial fisheries unprofitable without large government subsidies (Sala et al., 2018), while the rise of aquaculture has spurred technological innovations that continue to decrease production and environmental costs and increase yields, with significant potential for future growth (Gentry et al., 2017a). Aquaculture is currently >2.5 times more labor efficient (tonnes produced/worker) than fisheries and these farming efficiencies have increased by nearly 50% while fisheries have decreased by >32% since 1995 (Fig. 2a). Furthermore, 'fish in-fish out' ratios (a measure of feed conversion efficiency of a weight-equivalent of wild to cultured animals) have decreased by 65% for all aquaculture species (Fig. 2b), and some aquaculture has been shown to be more environmentally efficient than a suite of other animal protein sources (Poore and Nemecek, 2018). While wild fisheries require no feed at all, the relative efficiencies of fisheries and aquaculture production, coupled with increased demand and limited resources, signify that, like the Neolithic Revolution, transition to largely fish farming for seafood instead of wild capture is on the horizon for some regions. No doubt, timing of such a shift will continue to be heterogeneous, and will not result from one single event, similar to the transition of terrestrial food.

*Cultural diffusion*

There are seven main regions of agricultural origin, the earliest being traced to the Near East (Bellwood, 2005). While agriculture may have evolved independently in these major centers (Bellwood, 2005), mechanisms driving cultural diffusion of agriculture during the Neolithic Revolution fall under two primary hypotheses: 1) migration of immigrant farmers and/or 2) transmission of ideas through established trade and exchange networks (Larson et al., 2007). Interestingly, the rate of agricultural spread in different Neolithic populations seems to have been remarkably variable, with some populations exhibiting a rapid and complete change from hunting and gathering to agriculture after the first appearance of domesticates, while others adapted more slowly, or hardly at all (Bellwood, 2005; Richards et al., 2003; Richerson et al., 2001). However, agriculture eventually spread to impact and be adopted by nearly all human populations, and studies have shown that differences in timing of agricultural adoption can explain the nearly 100-fold differences in economic development across countries today (Hibbs and Olsson, 2004).

Currently, aquaculture also appears to have similar spatial and temporal variability in the rate of adoption (Fig. 3a). Mariculture– or farming in the marine environment – in particular has been shown to exhibit adoption patterns that are influenced by a number of factors, including taxonomic group and composition of production, economic development, and governance indicators within countries (Anderson et al., 2019; Gentry et al., 2019). China, for example, has a long history of aquaculture, dating back potentially ca. 8,000 years for common carp (Nakajima et al., 2019), and now, thanks to access to capital and political momentum (Davies et al., 2019), accounts for ca. 60% of global aquaculture production (freshwater and marine), with increasing production patterns and plans for continued growth (FAO, 2018; Gentry et al., 2019). Currently ca. 80% of China's total seafood production comes from aquaculture (the remaining 20% from fisheries - many with declining fish stocks and average trophic level of catches (Cao et al., 2017)), compared to 92% of production from fisheries in 1950 (Fig. 3b).

Conversely, in the United States, fisheries contributions remain relatively stable at >90% of total seafood production since 1950 (Fig. 3b). A letter from U.S. fishers opposing marine finfish aquaculture in U.S. waters captures the underlying sentiment that preserves this domestic fisheries dominance (excluding imports), referencing the dependent livelihoods and long-held value of sustainable commercial fishing ("America's oldest [industry]") in American society (American commercial fishers, 2018). Similarly, Chile, New Zealand, and British Columbia have all enacted moratoriums on certain types of aquaculture expansion at varying points in time, largely driven by social carrying capacity (i.e. avoiding adverse impacts on total human use of ocean space) and legislative transition (Banta et al., 2009; Bjørndal and Aarland, 1999; Noakes, 2003).

Although aquaculture has been practiced for thousands of years in some locations (Costa-Pierce, 2002; Nakajima et al., 2019), the impetus and potential for a more ubiquitous, global paradigm shift to aquaculture production was largely unthinkable until recently, likely due to social (e.g. NIMBY-ism), economic (e.g., governance, business friendly environments) and environmental (e.g., fisheries resources availability, sustainability of production) factors that have contributed to the drastic spatial and temporal variation in aquaculture innovation between countries. Perhaps aquaculture growth is at a point where cultural ties to fisheries are just beginning to be outweighed by demand (Fig 3c), capital gain, and relative efficiencies of aquaculture (Fig. 2a), similar to several of the hypothesized feedbacks experienced by our Neolithic ancestors.

*Property rights*

The transitory nature of mobile species and increasing scarcity of natural resources made sedentary lifestyles of our early ancestors difficult, but also led to the desire and ability to own and domesticate food (Weisdorf, 2005). Property rights are thought to have co-evolved with farming practices, giving humans the ability to live outside natural biological constraints and increase their resource base for the first time in history (Bowles and Choi, 2013; North and Thomas, 1977). Farming was initially less productive than foraging, but it has been suggested that exclusive ownership helped to spur significant increases in farming productivity (Bowles and Choi, 2013; North and Thomas, 1977). Even today, land rights have a strong relationship with agricultural yields and efficiencies (Chari et al., 2017; Goldstein and Udry, 2008).

Property rights have played a similar role in the evolution of fisheries, with strong ties to production levels and economic growth (Anderson et al., 2019). The sea was historically viewed as a common resource, with "property rights" not grounded in formal management. As competition and conflict grew (Spijkers et al., 2019) through technological innovation to "win the chase for the fish" (Anderson, 2002), so did demand for regulation and user rights. In 1982, the United Nations Convention on the Law of the Sea recognized the growing economic pressures over ocean resources and established sovereign rights to nation states over their adjacent waters (i.e. "Exclusive Economic Zones" (United Nations, 1982)). Differences in accessibility between land and sea has required fisheries management to evolve various forms of "user rights" including gear-type bans, catch restrictions, individual transferable quotas (ITQs), and territorial use rights for fishing (TURFs), all with varying levels of success (SAPEA Science Advice for Policy by European Academies, 2017). These legal instruments can greatly improve fisheries efficiencies (i.e., catch per unit effort and long-term availability of fished stocks) but numerous levels of security, durability, transferability, and exclusivity create complexities that make achieving full efficiency difficult (Arnason, 2012; FAO, 2014). Additionally, the many stakeholders vying for ocean space and the elusive balance between recreational, commercial and conservation interests commonly give rise to conflicts that property rights instruments, like ITQs, have not been sufficient in harmonizing (Arnason, 2012).

As fisheries resources become scarcer – through depletion, distributional shifts (*see* below), or simply limited supply – incentive for increased control and ownership continues. A transition to aquaculture provides a means for increased control over production on such a continuum (Anderson et al., 2019). However, in many areas, necessary frameworks for rights in the sea have not been established, providing a major roadblock for aquaculture development in some regions, but not others. In fact, studies investigating social acceptability of ocean-based industries (i.e., "social license to operate") have found the intangible impacts of maritime industries, largely relating to the values, beliefs and ideologies of a community, are harder to manage and are often left to decision makers in the political realm (Voyer and van Leeuwen, 2018). In New Zealand, for example, gaining rights for aquaculture production has been challenging for marine farmers, largely due to social factors and the apparent priority of other resource uses (Banta et al., 2009). Additionally, since traditional fisheries lack direct rights to ocean space, some fisher groups are concerned that property rights given to aquaculture through leases and permits will limit access to fishing grounds or displace fisheries catch (American commercial fishers, 2018), but it has been shown that aquaculture has thus far supplemented, not displaced, fisheries catches at the global scale (Longo and York, 2019). The question then becomes whether social license can stop, or only temporary forestall, production of aquaculture in a given region given the other drivers of change.

*Climate change*

The complexities of property rights coupled with concerns of overpopulation and overexploitation are being compounded by climate change, a phenomenon that may have also contributed to the invention of agriculture. The role of climate change in the Neolithic Revolution is still contested (Maher et al., 2011), however, several theories have been proposed such as climate change contracting communities into close association with animals in oases(Childe, 1928) (now largely discredited), increasingly stable climates spurring agricultural development (Richerson et al., 2001), and/or a shift to colder and dryer climates (the "Younger

Dryas") that may have decreased the abundance of wild cereals and motivated cultivation (Ashraf and Michalopoulos, 2011). Unlike the Younger Dryas, modern climate change is anthropogenically driven and occurring at rates never before seen in history (Karl and Trenberth, 2003), which is significantly impacting oceans by causing shifts in species distributions, sea level rise, mass coral bleaching events, decreased ocean productivity, altered food web dynamics, and greater incidence of disease, among other impacts (Hoegh-Guldberg and Bruno, 2010).

Globally, climate-induced changes in geographical distribution of fisheries production have been detected (Cheung et al., 2013), and are predicted to increase in the future (Barange et al., 2018). These shifts alter who has access to which stocks, and their expected overall productivity (Gaines et al., 2018). Ocean warming is estimated to have already decreased maximum sustainable yield of some populations by a combined 4.1% from 1930 to 2010 (Free et al., 2019). Climate change impacts on realized catch will largely depend on management measures, but there is evidence that climate change is already impacting some fisheries at a local scale. For example, fishermen in Oregon and California, USA sued big oil companies for warming-related algal blooms that have rendered the West Coast's Dungeness crab fishery unsafe for consumers, the first lawsuit of its kind (USGCRP, 2018). California's coast is also being hit by mass kelp die-offs and increases in purple sea urchin populations from an extreme El Nino event in 2014-2016, compromising red abalone populations and leading to a fishing closure for two seasons (Duggan, 2018). However, ocean warming will also open new ice-free areas that could provide new production potential for both aquaculture and fisheries (Christiansen et al., 2014; Froehlich et al., 2018).

Aquaculture's ability to adapt to wider environmental variability through genetic selection, relatively greater location flexibility (e.g., offshore, deeper waters, closed systems), and technological innovations have led some researchers to suggest aquaculture as a prospective mitigation strategy to future climate induced fisheries declines (Dey et al., 2016; Nwosu et al., 2016), providing further impetus for an eventual shift away from fishing to farming. Some former lobstermen in Maine, USA have even switched to farming oysters to diversify their income in a time of climate uncertainty (Bidgood, 2017), and aquaculture has shown to help diversify global food systems more generally in the face of resource scarcity and climate change (Troell et al., 2014). However, aquaculture will still be impacted by climate change through changes in ocean productivity and circulation patterns, ocean acidification, sea level rise, increases in ocean temperatures and extreme climatic events, and constraints and competition for limited resources (e.g., water, food, space, etc.), which may compromise or shift production potential (Barange et al., 2018). In fact, nearly all countries are expected to experience some net loss in marine aquaculture production potential in the coming decades, which may challenge aquaculture's capacity for growth in some regions (Froehlich et al., 2018).

**Shifted baseline of the Blue Revolution**

Comparing drivers of the Neolithic and Blue Revolutions provides unique insight into the future of seafood. Similarities in major drivers and feedbacks perhaps suggest the potential for an end of commercial, wild-caught fisheries as a primary food resource in the future. However, there are several differences that create a distinct baseline for aquaculture development and large uncertainty around the future balance of fisheries and aquaculture in seafood supply. Below we detail three primary differences impacting the trajectory, rate of change, and/or potential replacement of fisheries by aquaculture in society: 1) technological and scientific advancements, 2) environmental awareness and collective action, and 3) globalization and trade.

*Technological and scientific advancements*

Significant scientific advancements since the Neolithic Revolution, especially in the last 100 years, have led to a better understanding of numerous processes and interactions governing our planet, and thus ourselves, and have allowed for unprecedented technological innovations that have increased our ability to adapt and learn. Modern technology, and the compounding of knowledge gained during previous agricultural eras, have already led to a dramatic difference in the pace of change between the Blue and Neolithic Revolutions. Increases in aquaculture efficiencies and yields have occurred in an historic blink-of-an-eye (Fig. 2a) compared to the millennia it took to reach similar levels in agriculture, largely due to similarities in biotechnology across both food sectors (Subasinghe et al., 2003). For example, control and improvement of terrestrial mammals for agriculture likely took hundreds of years but has been accomplished in mere decades for some fish species, often with four to five times higher genetic gains than breeding programs for land species (Teletchea, 2016).

Advances in technology have compensated for fish scarcity, making some commercial fisheries largely dependent on aquaculture through hatcheries programs that enhance wild stocks. For example, in any given year, up to 90% of wild-caught Pacific Northwest salmon start their lives in aquaculture hatcheries, a technique that, while somewhat controversial, has been used for decades and has significantly enhanced commercial fisheries (NOAA, 2017). Similar techniques can be seen in terrestrial, small-scale recreational hunting operations, but stocking within agriculture at a commercial scale would be highly inefficient, perhaps foreshadowing the future of seafood.

On the other hand, much of current aquaculture production is still highly linked to wild fisheries for feed and seed (Clavelle et al., 2019). Fed aquaculture species (i.e., many finfish and crustaceans) rely on wild forage fish fisheries (e.g., sardines, herrings, anchovies) as a main feed ingredient. Further, for many species and production types, it is still more efficient to source wild seed for aquaculture grow-out as opposed to complete domestication (i.e., closed-cycle production systems), a major difference from modern agriculture. Only production that is fully domesticated (i.e., does not require wild seed), unfed (e.g., filter feeders), and/or fed alternative sources (e.g., soy, maize, rice, algae, insects) of protein and oil (Hua et al., 2019; Tacon, 2020), are decoupled from fisheries production, which is a growing proportion of species (Teletchea, 2015). Interestingly, aquaculture production may be approaching a stage where future technological innovations seemingly increase some fisheries' reliance on aquaculture – for restoration, recovery or increased yields (e.g., hatcheries programs) – but reduce aquaculture's reliance on fisheries (e.g., closed-loop domestication and reduced wild fish requirements in feed, Fig. 2b), a positive feedback that could contribute to the potential for a paradigm shift in food production, particularly at local scales.

If seafood demand persists and fisheries resources remain inconsistent and/or scarce, growing commercial interests in aquaculture will continue to spur innovation, such as an increase in offshore aquaculture, better disease management, changes in feed composition, efficiencies and delivery methods, hatchery systems, production types and genetic selection, all of which could continue to increase yield and reduce environmental impacts (Klinger and Naylor, 2012). Importantly, these innovations will likely result in increased aquaculture production that could theoretically price out wild caught products. In fisheries, technological improvements are unlikely to garner more fish or lower prices. Currently the only ways to feasibly reduce prices for wild caught seafood are 1) more sustainable management, which is costly to implement and provides limited improvement even under the most optimistic scenarios (Costello et al., 2016), 2) even larger government subsidies that may have negative consequences for sustainability and trade (Costello et al., 2020b), or 3) development of mesopelagic fisheries (Alvheim et al., 2020), though such catch maybe more suitable as feed and the ecological impact of such exploitation is highly uncertain (Hidalgo and Browman, 2019). Regarding subsidies, if capacity-building subsidies end, a task actively being pursued by the World Trade Organisation (Costello et al., 2020a; Sumaila et al., 2019), some fisheries could be priced out of the market even sooner by similar but cheaper aquaculture or fisheries products. However, as technological innovations are often driven primarily by profits, it will be important to ensure aquaculture intensification does not result in the same environmental costs as in agriculture, particularly given the diversity of species and production types across aquaculture today (Klinger and Naylor, 2012; Ottinger et al., 2016).

*Environmental awareness and collective action*

Both commercial agriculture and fisheries production intensified during a time when the intrinsic value of natural environments was not necessarily fully recognized by all peoples. In hindsight, and especially post-Industrial Revolution, society has come to understand the vast impacts that agricultural expansion has had on our environment (Tilman, 1999; Venter et al., 2016) and the risks posed for surpassing ecological limits at local to global scales (Springmann et al., 2018). However, these impacts have also significantly contributed to our understanding of the many linkages (e.g., food webs, land and sea interactions) and potential problems associated with food production today (Cottrell et al., 2017). The resultant scientific knowledge coupled with higher levels of education in the public at large, has led to an increase in environmental awareness. This awareness can be a powerful force of change and is dramatically different in the modern era thanks to technology (e.g., internet, cell phones, social media) that has increased our capacity for communication, organization, and collective action across scales.

For better or worse, negative environmental impacts of food production have often led to technological innovations and development of policy and management designed to avoid further degradation. For example, while much of Earth's terrestrial surface is too modified by agriculture to ever return to its "original state," the importance of sustainable agricultural practices is now a fundamental policy goal for a number of national and international institutions (e.g., United Nation's Sustainable Development Goal 2.4 (United Nations, 2015)), and many individual farmers choose to adopt practices to reduce soil loss, minimize run-off, and improve the conservation value of their land (Knowler and Bradshaw, 2007). Similarly, the idea that many fisheries resources were limitless contributed to unchecked and/or poorly managed fisheries and severe depletion of wild stocks worldwide. However, effective retroactive regulation has been shown to reverse these trends. The United States is a prime example where fisheries began to recover after the Magnuson-Stevens Fishery Conservation and Management

Act required all federally managed fisheries to have annual catch limits (Sewell and Atkinson, 2013).

Unfortunately, despite these advancements, sustainable scaling of food production remains a challenge for many food sectors today, such as pasture expansion in the Brazilian Cerrado (IPCC, 2019) and shrimp aquaculture production throughout South-East Asia in the 1980s (Arquitt et al., 2005). Whether related to agriculture, aquaculture or fisheries, environmental degradation can occur because of poor management, often resulting in negative environmental impacts such as disease outbreaks (Lafferty et al., 2015), propagation of non-native species (Naylor et al., 2001) and pollution (Verdegem, 2013). As in agriculture and fisheries, negative environmental impacts of aquaculture have been considerably minimized or reduced as environmental awareness has grown and better technologies have been developed. For example, aquaculture feed conversion ratios (a proxy for environmental efficiency) for salmon have decreased from ca. 2.8 to ca. 1.2 over the past 30 years (FAO, 2017), and policies and regulations have been enacted to restrict the once widespread aquaculture-induced mangrove deforestation in South-East Asia (Hishamunda et al., 2009).

The promise and sustainability of aquaculture growth remains hopeful, but it is still very uncertain and will likely remain highly spatially heterogeneous due to economic, social and political factors that may prove difficult to integrate and manage (Figure 1). Furthermore, in many areas there is a generally negative public perception surrounding marine aquaculture, which impacts development and growth. There are risks of negative environmental and social impacts from aquaculture, however negative perceptions may also be influenced by lack of knowledge or local environmental impacts not directly related to aquaculture at all (Froehlich et al., 2017). Even so, negative public perception remains an important factor for the future of aquaculture expansion, and given the power of collective action, could impede a more ubiquitous adoption of aquatic farming globally.

*Globalization and trade*

Perhaps the largest and most important difference between the Blue and Neolithic Revolutions is globalization and international trade. Globalization, international trade and technological innovations have removed local barriers to aquaculture supply and have amplified the rate and impact of cultural diffusion in today's society (FAO, 2018). Food, for example, is an important part of culture but has drastically increased diversity and availability at local scales, shifting from distinct national cuisines to a melting pot of techniques and ingredients (e.g., the U.S. sushi craze (Bestor, 2004) and Westernization of Asian diets (Pingali, 2007)). This globalization is likely a strong contributor to the drastic increase in per capita seafood consumption from 9 to 20.3 kg over the past 60 years (FAO, 2018) that has driven aquaculture growth.

Globalization and international trade have enabled aquaculture to remain one of the fastest growing food sectors (by volume) in the world despite significant heterogeneity in aquaculture adoption. Local opposition to aquaculture still exists and is strongest in the Westernized world, where environmental, social and food safety concerns are proportionally greater, as reflected by negative press (Bacher, 2015; Froehlich et al., 2017). However, commercial fisheries alone no longer meet seafood demand (FAO, 2018) (Fig. 3c) and global trade and associated markets have removed regional limitations of production and promoted market expansion of aquaculture, and seafood in general. Even if countries take a firm stance of limited domestic aquaculture growth, imports of top consumed species still support aquaculture growth (Lewison et al., 2019). For example, the U.S. seafood market currently depends on exporting its largely sustainably fished products and importing more reasonably priced, but in some cases less regulated, farmed seafood (e.g., farmed salmon, shrimp, and tilapia) (Asche et al., 2016). Of note, even if all U.S. seafood (wild and farmed) were not exported, the country would still be approximately one million tonnes shy, and growing, of meeting current demand (FAO, 2013) (Fig. 3c).

**The future of seafood: the balance between commercial fisheries and aquaculture**

There are similarities between the catalysts of commercial agriculture and aquaculture, perhaps foreshadowing a transition away from fishing to aquatic farming over coming decades to centuries. We present two future, exploratory scenarios of seafood production: 1) a ubiquitous aquaculture transition or 2) commercial aquaculture and fisheries coexistence. These two science-informed narratives contain two sub-narratives that differ based on management and regulatory strategies (Figure 1).

*Aquaculture transition*

If a path reminiscent of land-based food systems is assumed, based on common drivers discussed above, aquatic food production in some countries might progress away from wild fisheries toward a focus on aquatic farming (Fig. 1a and d). There is currently little evidence of aquaculture being the primary driver displacing or reducing fishing effort (Longo and York, 2019; Nahuelhual et al., 2019), likely due to economies of scale. Wild capture fisheries will likely never be completely eliminated, but fisheries may be relegated to solely support aquaculture production (i.e., feed and seed sources), along with subsistence and recreational purposes (as is the case with hunting on land today). In many areas, recreational fisheries landings already rival or exceed commercial catch, with five times more recreational fisheries than commercial fisheries globally, and often no limits on effort or catch (Arlinghaus et al., 2019). Human advancements since the Neolithic Revolution, as discussed above, will likely accelerate the transitional process of aquatic farming-centric nations (e.g., Fig. 3b below 0.5 dashed line) in the coming decades to centuries, and there is already some evidence that this may be occurring in places like China (Zou and Huang, 2015). In fact, China is promoting and investing in offshore aquaculture and has committed to reducing total marine catch volume by 5 MMT, which, along with other policies, will facilitate a shift to aquaculture production (Cao et al., 2017; Szuwalski et al., 2020; Yu and Han, 2020).

Such a transition today could have serious consequences for aquatic environments, particularly if the absence and/or inefficiencies of aquaculture policy in many countries are not amended. If aquaculture production is left to grow unchecked, without necessary frameworks and previsions (scenario 1, Fig. 1a), our food production history on land may be more likely to repeat itself in the sea; coastal ocean spaces could become heavily degraded, overrun with large-scale, monoculture farms and relatively devoid of wildlife. This potential future may be even more likely if wild capture fisheries go unregulated and collapse fish stocks or seafood demand continues to grow, putting further pressure on aquaculture to quickly expand and fill a gap in supply. The aquaculture sector may already be headed towards farming relatively few aquatic species, like on land, considering >50% of total global aquaculture is currently produced by just 10 species (Ottinger et al., 2016), similar to the 10 crop species making up 70% of global production on land (FAO, 2013).

Perverse impacts on our environment from aquaculture expansion are not inevitable, and certainly not a desired outcome for our oceans or freshwater systems. Impacts on aquatic species and habitats can be accounted for and managed (scenario 2, Fig. 1d). If proactive, science-informed policies and regulations are implemented to avoid these environmental consequences before they occur, a better future of aquatic food production can be achieved. For example, rigorous spatial planning and standardized practices can help maximize aquaculture yields while minimizing environmental impacts and avoiding conflict with other ocean uses (Gentry et al., 2017b; S. E. Lester et al., 2018). In this scenario, a transition from commercial fishing to sustainable aquaculture could potentially leave millions of tons of wildlife in the sea by significantly reducing one of the most pervasive immediate threats to marine biodiversity, fishing (Halpern et al., 2010; Maxwell et al., 2016), while maintaining consistent seafood supply. Furthermore, if production is well-sited and managed, both production area and potential impacts may be a fraction of what is currently seen in wild caught fisheries and in agricultural production (Gentry et al., 2017a).

Different levels of legislation and policies do exist for aquaculture around the world, but can be limited, absent or lacking specificity, especially around production expansion (Asche et al., 2016; Davies et al., 2019). For example, offshore aquaculture (i.e., within a country's Exclusive Economic Zone) is coming online in countries around the world – including China and Norway, two of the largest finfish producers – but regulatory standards for offshore production are unclear (Davies et al., 2019; Sarah E. Lester et al., 2018). Not only has this been a roadblock for aquaculture expansion, it has often caused further tension between aquaculture and fisheries sectors. This tension is particularly evident in countries that lack clear national standards for aquaculture production, creating confusion surrounding bodies governing aquaculture and making it difficult for both regional policy makers and producers to navigate and manage (e.g., environmental regulation, permitting).

*Aquaculture and fisheries coexistence*

Until recently, fisheries held the long-standing dominant position in seafood production. The rise of aquaculture over the past 60 years has led to the increased coexistence of these two sectors at the commercial scale. However, in many cases commercial aquaculture and fisheries are not co-developed or managed and there are not specific policies to ensure integration, efficiency, or even the continued coexistence of both sectors into the future (e.g., a transition to aquaculture). The degree to which aquaculture and fisheries coexist cooperatively or competitively at regional to global scales will continue to have direct impacts on both industries and the environment, and will depend on how rights to aquatic space are managed and the careful development of policies that recognize the highly integrated nature of seafood production (Klinger et al., 2017).

As we previously discussed, coexistence of aquaculture and fisheries sectors is pervasive in many countries today (scenario 3, Figure 1b). However, uncoordinated coexistence can lead to inefficiencies in production because the integrated nature of aquaculture and fisheries are not recognized (e.g., aquaculture feed ingredients and hatcheries contribution to fisheries), can put further pressure on finite resources and the environment as competition for ocean resources increases, and can have significant implications for each sector as new seafood supply is added or removed from the market space (Clavelle et al., 2019). Without clear frameworks, this competition may be more likely to manifest through campaigns that block legislative measures to expand aquaculture or manage fisheries/aquaculture conflicts, particularly when operations are difficult to separate (e.g., capture-based aquaculture and culture-based fisheries) (FAO, 2018).

Given the inherent linkages of these two sectors (Clavelle et al., 2019), better outcomes will be achieved if synergies are exploited through integrated policy and systematic planning (scenario 4, Fig. 1c). To promote a more holistic seafood system, aquaculture could be integrated into the comparatively more well-defined structure of ecosystem-based fisheries management (EBFM) frameworks (Klinger et al., 2017). An integrated aquaculture framework may include, but is not limited to, guidelines and regulations surrounding clear regulatory structure, science-based reference points (e.g., carrying capacity), agreeable targets of scaled production (reverse-quotas, if you will, on quantity, extent, and/or production types), spatial extent of overlap or interaction (e.g., avoid critical habitat and productive fishing grounds), access and use of wild seed for culturing species (e.g., shellfish), identifying 'hatchery' as an aquaculture contribution to fisheries, and/or implementing 'trade caps' on imported farmed seafood products, supplemented by domestic aquaculture production. Clearly defining aquaculture's interactions and role in maintaining fisheries, and vice versa, we argue, will set up greater success for both industries and the environment. This may be particularly true when population and seafood demand plateau and economic and social considerations overtake the underlying need to supply demand.

Better communication and investigation of real versus perceived interactions between aquaculture and fisheries are also needed so informed decisions can be made by stakeholders and decision makers. Such policy can only be developed by the equal inclusion of all relevant stakeholders (which can be difficult (Voyer and van Leeuwen, 2018)) in planning decisions (SAPEA Science Advice for Policy by European Academies, 2017). Iceland recently took a step in this direction when the Icelandic Federation of Fish Farming joined as part of the Confederation of Icelandic Fishing Companies, the organization that represents deep sea trawling interests (McLeod, 2018). However, even if management and regulatory guidelines are properly implemented, aquatic food production may still transition to aquaculture (for the many reasons outlined above). Moreover, it is unlikely for fisheries to return to its place of dominance in the seafood sector.

The evidence base and narratives presented here largely refer to aquatic farming – comprised of both marine and freshwater aquaculture and, in fact, most of the spatial, economic, social and environmental interactions between commercial fisheries and aquaculture can and do apply to both sectors. However, many of the examples and referenced literature focus on marine seafood production due to the predominance of commercial marine catches in fisheries and the immense capacity for mariculture growth (Costello et al., 2020a). Freshwater aquaculture accounts for ~60% of current aquaculture production and will be an important component of seafood in the future. There is likely to be nuanced differences between freshwater aquaculture growth, freshwater fisheries and agriculture. Relationships between these three sectors and the patterns of a paradigm shift should be explored further through a freshwater lens, particularly as data continues to become available.

**Conclusions**

From the patterns and scenarios discussed above, it is clear that the balance between fisheries and aquaculture is not black and white. However, we argue that ignoring aquaculture's role in seafood production is no longer an option. If aquaculture and fisheries are to sustainably coexist locally, aquaculture development needs to be guided by both stand-alone and integrated policy and regulation, which is currently lacking and often heavily contested. Society now has the foresight and capacity to choose what the future should look like for global seafood production and ocean health through careful planning, policies, and management. Determining *a priori* objectives and developing regulations to achieve outlined goals, while minimizing social and environmental impacts, are urgently needed to increase the stability and longevity of both aquaculture and fisheries in the future by forcing articulation of expectations for a continually growing global seafood sector.

**Acknowledgements**

The National Center for Ecological Analysis and Synthesis at UC Santa Barbara provided invaluable infrastructural support for this work. CDK, HEF and BSH acknowledge funding from the Zegar Family Foundation. We thank Allison Horst for her collaboration and artistic contribution on Table 1 and Figure 1. HEF is a member of the Technical Advisory Group for the Aquaculture Stewardship Council.

Hawkes, C., Zurayk, R., Rivera, J.A., De Vries, W., Majele Sibanda, L., Afshin, A., Chaudhary, A., Herrero, M., Agustina, R., Branca, F., Lartey, A., Fan, S., Crona, B., Fox, E., Bignet, V., Troell, M., Lindahl, T., Singh, S., Cornell, S.E., Srinath Reddy, K., Narain, S., Nishtar, S., Murray, C.J.L., 2019. Food in the Anthropocene: the EAT–Lancet Commission on healthy diets from sustainable food systems. The Lancet 393, 447–492. https://doi.org/10.1016/S0140-6736(18)31788-4

Yu, J., Han, Q., 2020. Food security of mariculture in China: Evolution, future potential and policy. Marine Policy 115, 103892. https://doi.org/10.1016/j.marpol.2020.103892

Zou, L., Huang, S., 2015. Chinese aquaculture in light of green growth. Aquaculture Reports 2, 46–49. https://doi.org/10.1016/j.aqrep.2015.07.001
34

**Table 1. Comparison of the catalysts driving the Neolithic and Blue Revolutions.** Population, cultural diffusion, property rights, and climate change during the Neolithic Revolution compared to trends in these catalysts today, which may be foreshadowing a transition from commercial fishing to aquatic farming. Arrows depict key differences in today's world and whether trends suggest they are pushing seafood production towards fishing (environmental awareness and collection action) or towards fish farming (technological and scientific advancement, globalization and international trade, environmental awareness and collective action). Illustrations by Allison Horst.

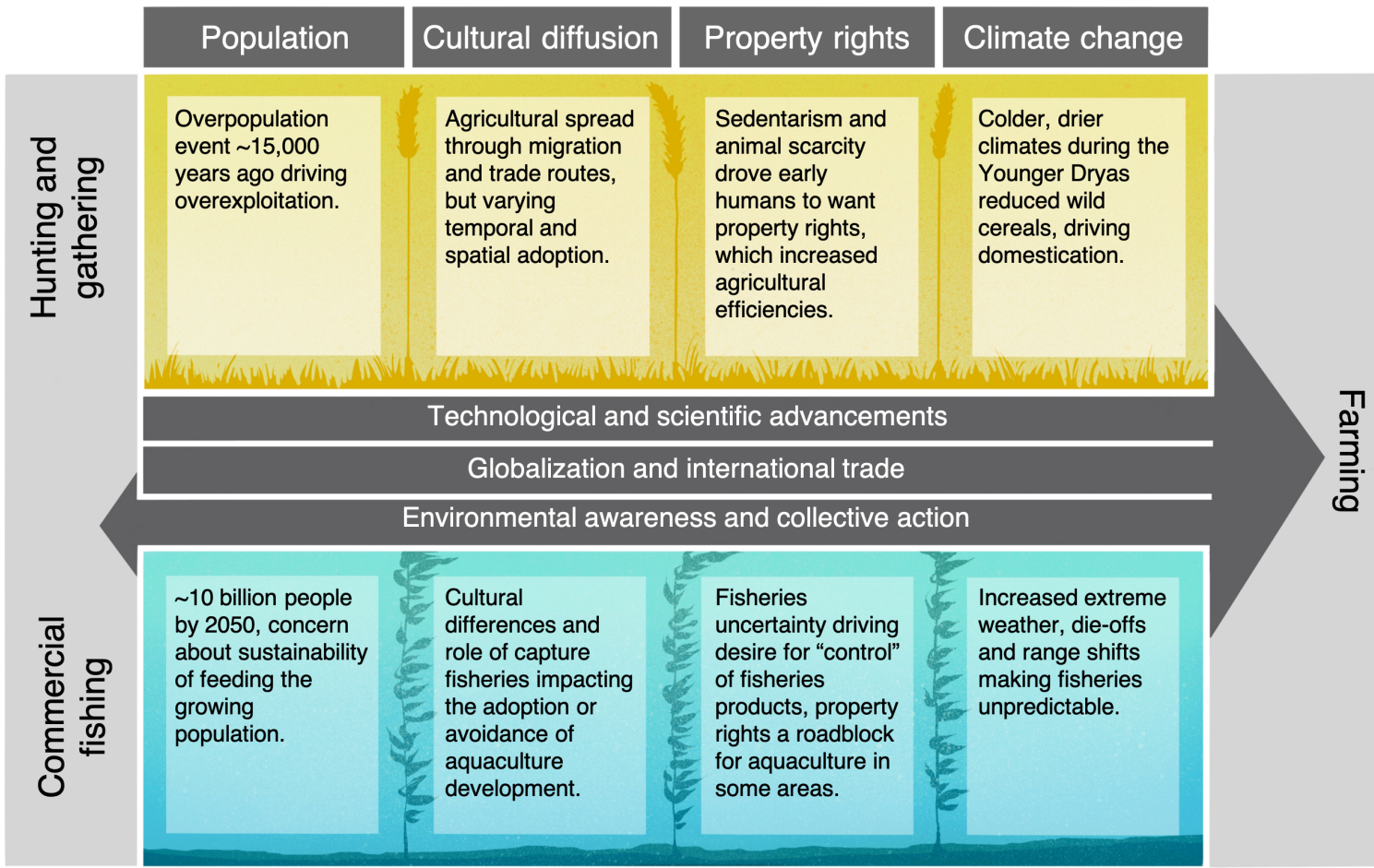

**Figure 1. Varying balances between commercial aquaculture and fisheries in future seafood production and the potential impacts on our oceans. (a)** Seafood production transitioning away from fishing to aquatic farming due to economic pressures and relative efficiencies, and resulting in rapid habitat conversion and loss in biodiversity from unsustainable practices and unregulated growth**. (b)** Commercial fisheries and aquaculture both continue to contribute to global seafood supply but competition due to lack of regulation and guidelines results in a crowded ocean, with potential for environmental degradation and hostility. **(c)** Aquaculture and fisheries sustainably coexist to meet the rising demand for seafood while minimizing conflict and environmental degradation through integrated and effective management. **(d)** Environmental and economic pressures lead to an aquaculture transition, but holistic management strategies and controlled, carefully planned growth avoid the pitfalls of agriculture on land while maintaining seafood supply. These scenarios highlight just four of the numerous potential habitat, production and cooperation states that may occur given the complex nature of the seafood sector. Illustrations by Allison Horst.

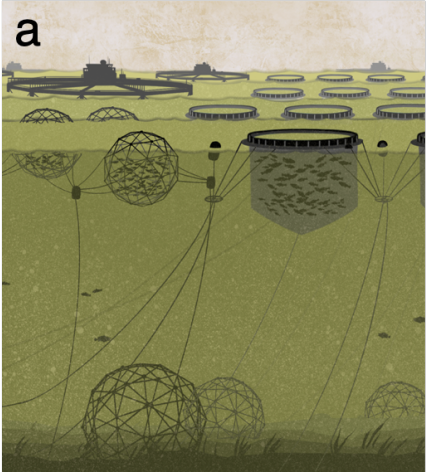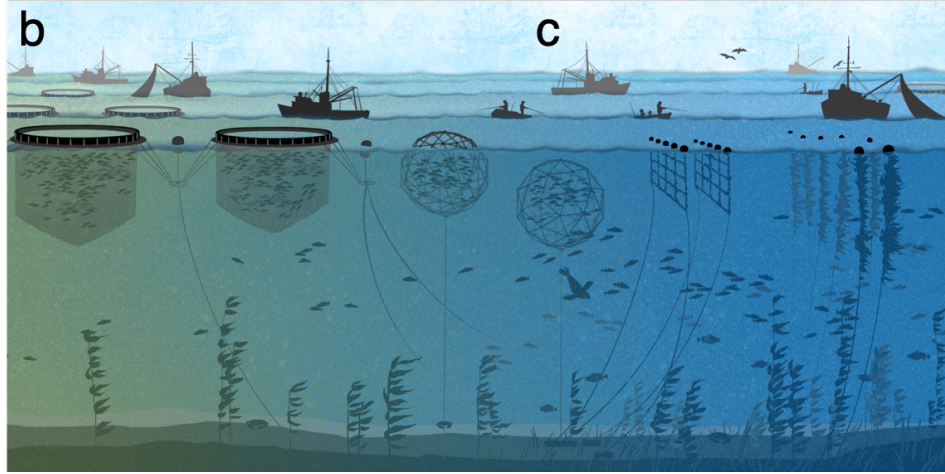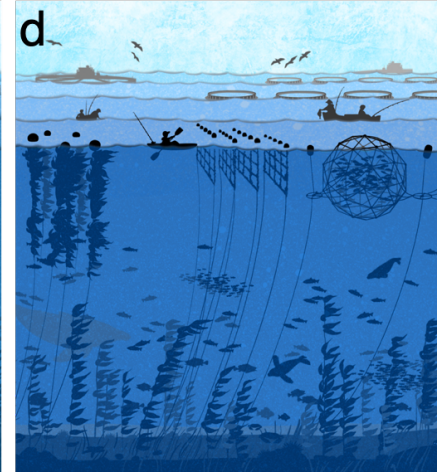

**Figure 2. The relative efficiency of aquaculture and capture fisheries. a)** The respective global tonnes of aquaculture or commercial fisheries production per farmer or fisher. Through time (5-year increments from 1995 to 2010, 1-year increments thereafter, no data for 2011) and **b)** the fish in-fish out ratio (the amount of wild fish products used to produce one weight equivalent of farmed fish) for various aquaculture species groups in 2000, 2010, and 2015. Production and employment data from FAO(FAO, 2018). Fish in-fish out ratios were sourced from The Marine Ingredients Organization(The Marine Ingredients Organization, 2017). Silhouette images from phylopic.org and the IAN symbol library (http://ian.umces.edu/symbols/).

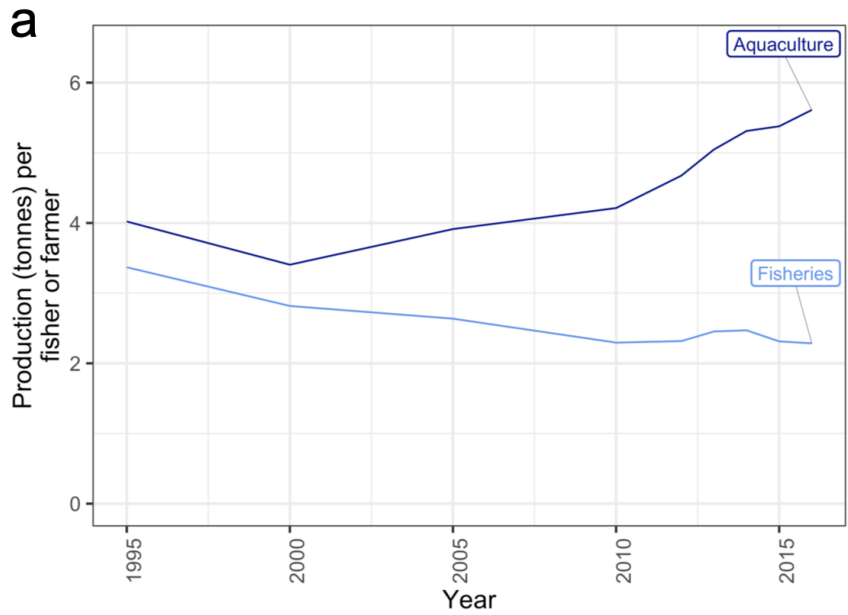
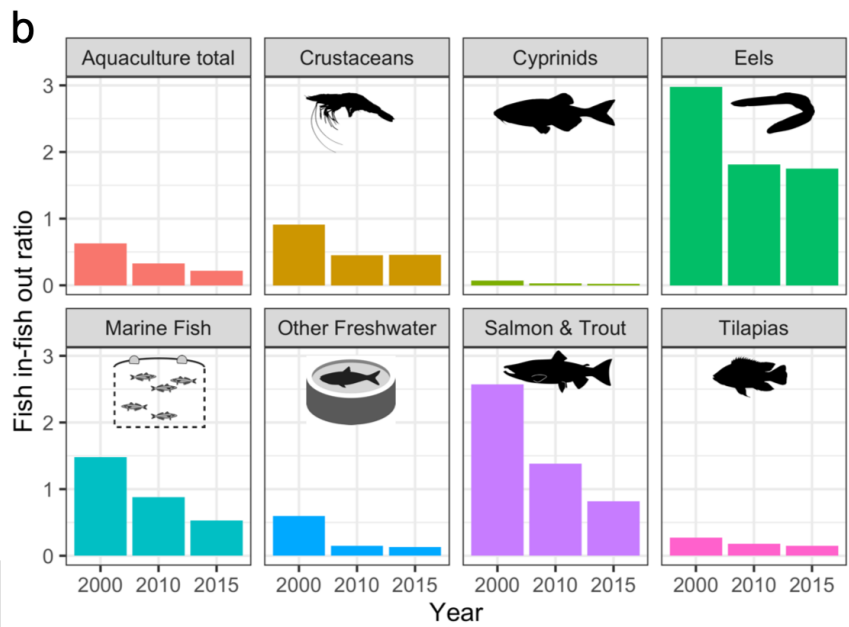

**Figure 3. Trends in aquatic food production and consumption through time of the top 10 aquatic food producers. a)** The proportion of aquatic food production from capture fisheries by country in 2016 and **b)** from 1960 to 2015 for the top 10 aquatic food producers. A value of 1 indicates that countries produce all aquatic food through fisheries, a value of 0.5 indicates that aquaculture and fisheries contribute equally to aquatic food production and a value of 0 indicates that all aquatic food is produced through aquaculture. **c)** The ratio of aquatic food consumption vs. production through time. Countries that consume more than they produce have a value >1 while countries that produce more than they consume have a value <1. Production and consumption data can be downloaded from FAO(FAO, 2013). Trends depict 5-year moving averages calculated with the zoo package in R v3.5.1.

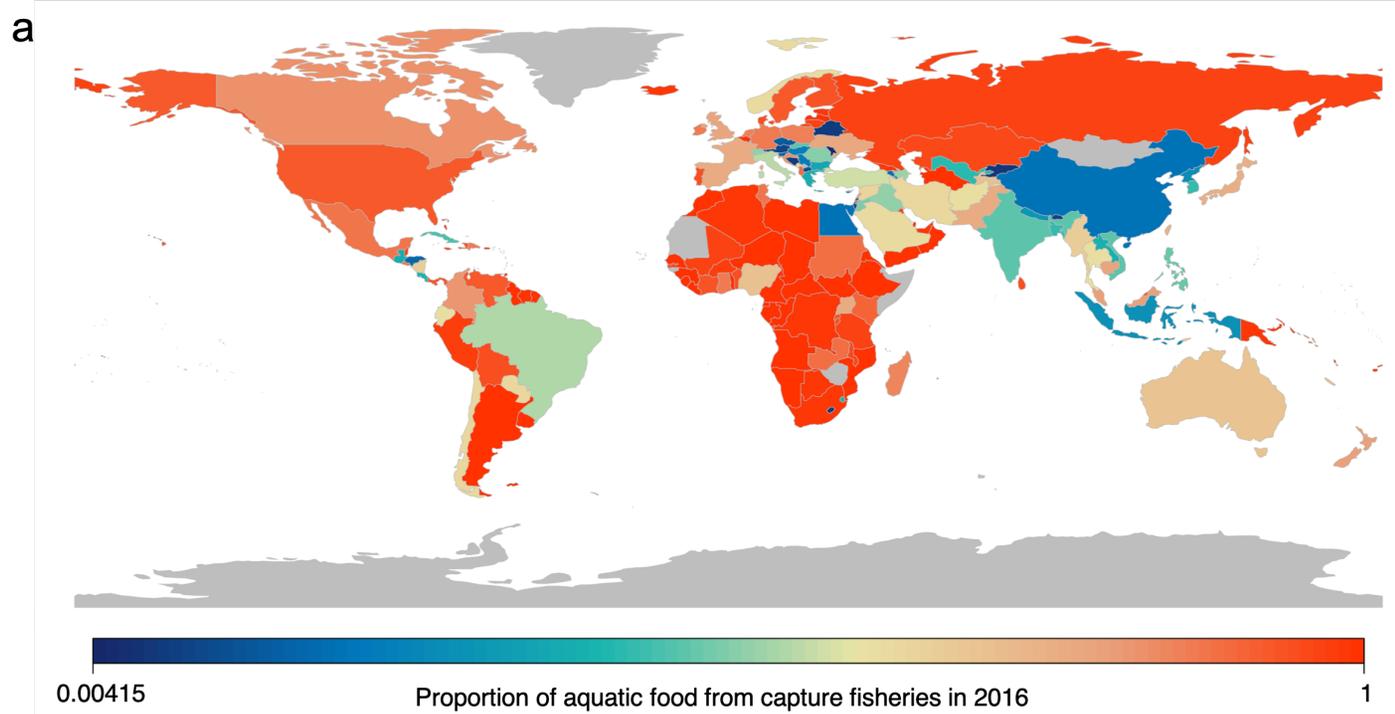

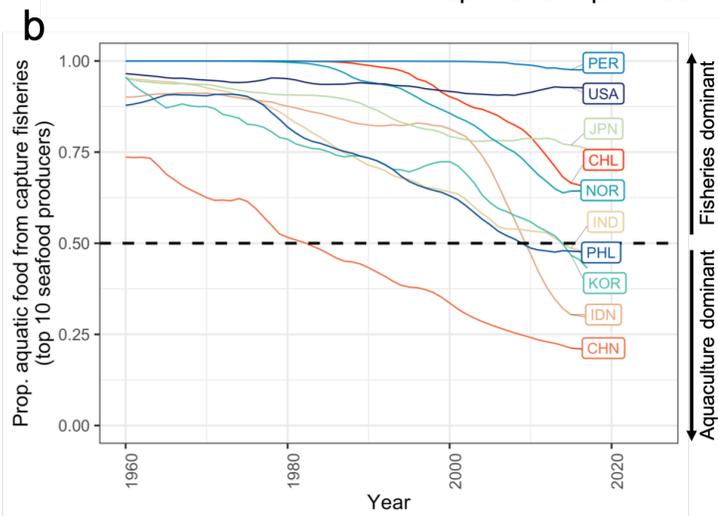

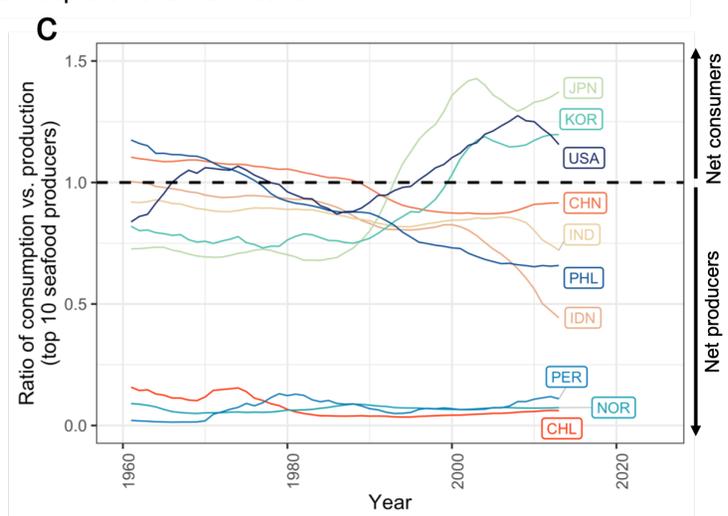